\documentstyle{elsart}

\begin{document}
\begin{frontmatter}

\title{Theory of Ideal Four-Wave Mixing in Bose-Einstein Condensates}
\author[a,b]{ Wenji Deng \thanksref{phwjdeng}}, and
\author[b]{ P. M. Hui}
\address[a]{Department of Physics, South China University of Technology,\\
Guangzhou 510641, China.}
\address[b]{Department of Physics, The Chinese University of Hong Kong,\\
Shatin, New Territories, Hong Kong, China.}
\thanks[phwjdeng]{email address: phwjdeng@scut.edu.cn}

\maketitle

\begin{abstract}
Starting from a second-quantized Hamiltonian of many-particle systems, 
we derive a set of time evolution equations for four-wave mixing (4WM) 
processes of coherent matter waves, which is analogous to those 
in optical 4WM except that the spatial variable is being 
replaced by time.
Several interesting problems in 4WM such as 
the phase-matching condition, the effects of relative phase 
difference, and the conversion efficiency are then discussed in detail.
We also show that the main features in recent 
4WM experiment [Deng {\it et al}, Nature 398, 218 (1999)] can be 
undersood 
within the present simplified model.\\
PACS number(s): 03.75.Be, 03.75.Fi
\end{abstract}

\end{frontmatter}

\section{ Introduction}
The experimental realizations of Bose-Einstein condensation 
(BEC) in trapped dilute alkali atomic vapours \cite{exp} 
and the atom ``laser''\cite{laser} have made possible 
the experimental study\cite{deng} of four-wave mixing (4WM) 
of coherent matter-waves. 
As a typical effect of nonlinear atom optics\cite{lens}, 
the possibility of 4WM with matter waves of 
Bose-Einstein condensate (BEC) can be deduced from the 
similarity between the Maxwell equations for lasing action 
in nonlinear optical media\cite{optical} and 
the Gross-Pitaevskii(GP) equation\cite{GP} widely used 
in studying the properties 
of macroscopic quantum systems such as BECs.

In the past few years, 4WM in BECs and related problems 
have been studied intensively
\cite{deng,conjugation1,conjugation2,4wm1,spinor1,spinor2,spinor3,q4wm,4wm2,wu,4wm3}. 
Early in 1995, Goldstein {\it et al.} had started to 
investigate the problem of phase conjugation\cite{conjugation1} 
in BECs, including the case of multi-component BECs\cite{conjugation2}. 
A 4WM experiment using three colliding BEC wavepackets 
with different central momenta was proposed by 
Trippenbach {\it et al.}\cite{4wm1}.
Recently, Deng {\it et al.} successfully carried 
out 4WM experiment in sodium condensates\cite{deng} 
by means of Bragg diffraction technology\cite{Bragg}.
In addition, stimulated by the  
work of Ho and coworkers\cite{spinor1} 
and Ohmi and Machida\cite{spinor2} on the Bose condensates 
with internal degrees of freedom,
Law {\it et al.} studied quantum spin-mixing 
in spinor BECs\cite{spinor3} using 
an algebraic method developed in problems related to 
cavity QED\cite{algebraic},
and Goldstein and Meystre\cite{q4wm} developed 
a quantum theory of atomic 4WM involving both 
the internal and spatial degrees of freedom.
Rz\c{a}\.{z}eewski {\it et al.} investigated the fluctuations
in the populations of atoms in 4WM wavepackets due to the 
quantum mechanical nature of the mean-field BEC wavefunctions\cite{4wm2}.
Although the experimental results turn out 
to be consistent with the numerical simulations 
by Trippenbach {\it et al.}\cite{4wm1} based on the GP equation, 
further theoretical studies are still needed 
in order to understand  4WM in BECs more clearly.
Without invoking the undepleted pump approximation, 
Wu {\it et al.}\cite{wu} calculated analytically the 4WM in 
the experimental condition of Ref.\cite{deng}.
Based on the slowly-varying-envelope approximation, 
Trippenbach {\it et al.}\cite{4wm3}
developed a three-dimensional quantum mechanical description 
for 4WM of wavepackets created from BEC. 

The main purpose of the present work is to 
study the ideal 4WM with coherent matter waves from BEC 
in which only four plane waves with definite wavevectors 
and frequencies are involved. 
We organize the paper as follows.
A set of time evolution equations for 4WM are derived 
in Sec.\ref{sec:GP} starting from a Hamiltonian for 
many-particle systems in the second quantized form 
rather than the original GP equation. 
Section \ref{sec:pm} gives a short discussion on the 
phase-matching condition for ideal 4WM.
In Sec.\ref{sec:d}, we present results demonstrating the effects 
of relative phase difference and the collapse and revival (C\&R) 
behaviour of ideal 4WM.
The problem of conversion efficiency and the main characters of 
recent 4WM experiment\cite{deng} are investigated in Sec.\ref{sec:ce}.
Sec.\ref{sec:summary} gives a brief summary.

\section {Time evolution equations for 4WM}\label{sec:GP}

To study 4WM with matter waves, we consider 
the dynamics of $N$ interacting identical bosons. 
For simplicity, we omit the internal degrees of freedom. 
In standard experimental setups for studying 4WM, the confining 
potential is removed,
and the Hamiltonian can be generally written in terms of 
bosonic creation and 
annihilation operators as\cite{Fetter}
\begin{equation}
\label{eq:Hamiltonian}
  {\mathcal{H}}=\sum_{k}\varepsilon_k\hat{a}_k^{\dag}\hat{a}_k
  +\sum_{{\mathbf{k}}_{4}+{\mathbf{k}}_3={\mathbf{k}}_1+{\mathbf{k}}_2}
  \frac{U_q}{ {2\mathcal{V}} }
  \hat{a}_{k_4}^{\dag}\hat{a}_{k_3}^{\dag}\hat{a}_{k_2}\hat{a}_{k_1},
\end{equation}
where $\varepsilon_k=\hbar^2k^2/2M$ is the kinetic energy of a particle in
a single-particle state of momentum ${\mathbf{p}}=\hbar{\mathbf{k}}$,
${\mathbf{q}}\equiv {\mathbf{k}}_4-{\mathbf{k}}_1={\mathbf{k}}_{2}-{\mathbf{k}}_{3}$
denotes the momentum transfer in two-body scattering, 
and $M$ denotes the atomic mass.
Momentum conservation in the second term results from 
the calculation of the interaction
matrix elements  
$U_{{\mathbf{k}}_4{\mathbf{k}}_3{\mathbf{k}}_2{\mathbf{k}}_1}=U_{q}
\delta_{{\mathbf{k}}_4+{\mathbf{k}}_3,{\mathbf{k}}_1+{\mathbf{k}}_2}$,
where 
$U_{q}\equiv \int d^{3}{\mathbf{r}}U(r)\exp(i\mathbf{q}\cdot\mathbf{r})$
is the Fourier transform of the interaction potential $U(r)$.  
The field operator can be expanded as
$\hat {\Psi}({\mathbf{r}},t)=\sum_{k}\hat{a}_{k}
\exp(i{\mathbf{k}}\cdot{\mathbf{r}})/\sqrt{{\mathcal{V}}}$ 
with  ${\mathcal{V}}$ denoting the volume of the system. 

Within the Heisenberg picture, the wavefunction is independent of time 
and the dynamical evolution of the system is given by that of the field operators. 
The annihilation operators satisfy equations of the form 
\begin{equation}
\label{eq:Operator}
i\hbar\frac{\partial \hat{a}_{k_4}}{\partial t}=
( \varepsilon_{k_4}+U_0\hat{n})\hat{a}_{k_4}+
\sum_{\stackrel{\mathbf{k}_{1}+\mathbf{k}_{2}-\mathbf{k}_{3}=\mathbf{k}_{4}}
{q\ne 0}}
\frac{U_{q}}{\mathcal{V}}
\hat{a}_{k_3}^{\dag}\hat{a}_{k_2}\hat{a}_{k_1},
\end{equation}
where the subscripts $k_1, k_2, k_3$ and $k_4$ can be any 
allowed values of wavevectors (or particle momenta), and 
the operator of the averaged number density of atomic gas 
is a fixed constant defined by 
$\hat{n}\equiv (1/{\mathcal{V}})\sum_{k}\hat{a}_k^{\dag}\hat{a}_k$.

Guided by results obtained through the experimental realization 
of BEC and the
technique of Bragg diffraction\cite{deng,Bragg},
we assume that the creation and annihilation 
operators $\hat{a}_{k}^{\dag}$
and $\hat{a}_{k}$ of a certain mode with {\em macroscopic}
population of atoms can 
be treated as a pair of conjugated complex numbers $a_k^{\ast}$ and $a_k$, 
and those for the other modes vanished\cite{spinor2,Bogoliubov,Anderson}. 
Therefore the field operator becomes a macroscopic 
wavefunction satisfying 
the GP equation, except that $U_{q}$ may depend on $q$ for scatterings
with large momentum transfer\cite{GP,Note1}. 
Introducing a new variable $A_k\equiv a_k/\sqrt{\mathcal{V}}$ and 
assuming for simplicity a $q$-independent  
$U_{q}\approx 4\pi a_0 \hbar^2/M$ with $a_0$ denoting 
the s-wave scatteing length,
the slowly varying  envelope function of the field, 
$\tilde{A}_k\equiv A_{k}\exp[i(\varepsilon_k-nU_0)t/\hbar]$ 
with $n=\sum_ka_k^\ast a_k/{\mathcal{V}}$, satisfies
a set of coupled nonlinear equations
\begin{equation}
\label{eq:GP}
i\frac{\partial \tilde{A}_{k_4}}{\partial t}=\frac{2 a_0 h}{M}
\sum_{
\stackrel{{\mathbf{k}}_1+{\mathbf{k}}_2-{\mathbf{k}}_3={\mathbf{k}}_4}
{q\ne 0}}
\tilde{A}_{k_3}^{\ast}\tilde{A}_{k_2}\tilde{A}_{k_1}
e^{i\Delta \omega t}
\end{equation}
with $\Delta \omega=(\varepsilon_{k_4}+\varepsilon_{k_3}-
\varepsilon_{k_2}-\varepsilon_{k_1})/\hbar$.
The momentum conservation indicated in the summation is 
a consequence of the translational invariance of the system. 
In contrast,  collision processes with $\Delta \omega \ne 0$
are {\em not} forbidden in principle.  
As the system has a fixed number of atoms,
the averaged number density is a constant, i.e., 
$\sum_k |\tilde{A}_k|^2=n$.
We can then define the normalized field amplitudes as
$\xi_k=\tilde{A}_k/\sqrt{n}$, 
which satisfy the normalization condition $\sum_k|\xi_k|^2=1$, 
to describe the relative distribution of atoms more conveniently. 
The time evolution equations for these
normalized amplitues take on the more compact form of 
\begin{equation}
\label{eq:MWM}
i\frac{\partial \xi_{k_4}}{\partial \tau}=\frac{1}{2}
\sum_{
\stackrel{{\mathbf{k}}_1+{\mathbf{k}}_2-{\mathbf{k}}_3={\mathbf{k}}_4}
{q\ne 0}}
\xi_{k_3}^{\ast}\xi_{k_2}\xi_{k_1}e^{i\Omega \tau},
\end{equation}
where the re-scaled time $\tau\equiv t/T$ is dimensionless with 
the characteristic time 
$T\equiv M/4na_0h$ being inversely proportional to the number 
density, and the phase mismatch or `energy-loss' 
parameter is defined as $\Omega\equiv T\Delta\omega$. 
Before going into details, some important conclusions can be drawn 
on the common properties of 4WM with matter waves.
The number density of atoms can affect the time evolution 
of 4WM only by modifying the characteristic time $T$, implying that
4WM evolves faster in situations of higher number density of atoms.
In actual experimental situations\cite{deng}, 
the characteristic time is estimated to be about $3ms$ (see Sec.\ref{sec:ce}).

The general form of time evolution equations for 
multi-wave mixing (MWM) is a rather 
complicated set of coupled nonlinear equations because 
the subscripts $k_1, k_2, k_3$ and $k_4$ in Eq.(\ref{eq:MWM}) 
can take on any allowed values of wavevectors (or particle momenta).
However it can be simplified for an ideal 4WM process to 
\begin{equation}\label{eq:4WMa}
\frac{\partial \xi_1}{\partial \tau}=
-i\xi_2^{\ast}\xi_3\xi_4 e^{-i\Omega\tau} ,\;
 \frac{\partial \xi_2}{\partial \tau}=
-i\xi_1^{\ast}\xi_3\xi_4e^{-i\Omega\tau} ,
\end{equation}
\begin{equation}\label{eq:4WMb}
\frac{\partial \xi_3}{\partial \tau}=
-i\xi_4^{\ast}\xi_2\xi_1 e^{i\Omega\tau} ,\;
\frac{\partial \xi_4}{\partial \tau}=
-i\xi_3^{\ast}\xi_2\xi_1 e^{i\Omega\tau} .
\end{equation}
This set of equations is analogous to those in optical 4WM 
except that the spatial variable is 
replaced by the re-scaled time.
In what follows, we will study  4WM of matter waves 
based on the simplified model described 
by  Eqs.(\ref{eq:4WMa}) and (\ref{eq:4WMb}). 
All numerical calculations are carried out using 
the 4th order Runge-Kutta method.

\section{Phase-matching condition for 4WM}\label{sec:pm}
The recent 4WM experiment\cite{deng} shows that 
three colliding wavepackets with arbitrary central momenta 
cannot always create a fourth wavepacket,
and a possible 4WM process must satisfy the 
phase-matching condition, {\it i.e.} momentum-energy conservation. 
The Hamiltonian (Eq.(\ref{eq:Hamiltonian})) describes a system 
satisfying the conservation of total energy, momentum and particle-number.
Therefore the phase-matching condition is always understood as 
${\mathbf{k}}_4+{\mathbf{k}}_3={\mathbf{k}}_1+{\mathbf{k}}_2$ and 
$\varepsilon_{k_4}+\varepsilon_{k_3}=\varepsilon_{k_2}+\varepsilon_{k_1}$.
Note that from the dispersion relation of massive particles 
$\varepsilon_k=(\hbar k)^2/2M$, 
the phase mismatch parameter can be written as
$\Omega=({\mathbf{k}}_3-{\mathbf{k}}_2)\cdot
({\mathbf{k}}_3-{\mathbf{k}}_1)/8\pi na_0$. 
The phase-matching condition, {\it i.e.} $\Omega=0$, 
for matter waves can be shown graphically as in Fig.1.

It has also been emphasized, however, 
in the discussion after Eq.(\ref{eq:Hamiltonian}) that 
the scattering process of bosons {\em does not} require 
the conservation of energy. 
So the constraint of energy conservation deserves some more explanations.
In fact, numerical calculations show that 
the parameter $\Omega \equiv 
T \Delta\omega$ affects 4WM processes, but only 
large enough phase mismatch can suppress the creation of fourth wave. 
Figure 2 shows the population $N_{4}$
of the created wave as a function of 
$\tau$ for different values of $\Omega$.  
For small $\Omega$, $N_{4}(\tau)$ is similar to that 
of an ideal 4WM process satisfying the phase-matching condition exactly.  
The conversion efficiency is lower for higher values of 
$\Omega$, and no 4WM signal can be observed when $|\Omega| \geq \pi$.  

\section{ Dynamics of Ideal 4WM}\label{sec:d}
The time evolution of an ideal 4WM system 
can be described by Eqs.(\ref{eq:4WMa}) and (\ref{eq:4WMb}). 
This kind of process satisfies not only particle-number conservation but also the conditions 
\begin{equation}
 \frac{\partial|\xi_4|^2}{\partial \tau}=
 \frac{\partial|\xi_3|^2}{\partial \tau}=
 -\frac{\partial|\xi_2|^2}{\partial \tau}=
 -\frac{\partial|\xi_1|^2}{\partial \tau},
\end{equation}
implying that the two pumping waves must provide 
atoms to amplify the probing and created waves at the same rate. 
When one of the two pumping waves is exhausted, 
the another one will take on the role of the probing wave. 
Figure 3 shows the calculated results for 
typical collapse and revival (C\&R) 
behaviour of 4WM. 
Initially, the $N$ atoms are distributed in three states 
corresponding to the  
pumping wave 1, pumping wave 2 and  probing wave 3.
Due to two-body interaction and the bosonic enhancement effect, 
the atoms in the pumping waves begin to transfer 
into the states 3 and 4.  
After a short period of time, all the atoms are transferred out of
state 2 and no more atoms are transferred into state 4. 
The second phrase of 4WM begins at around $\tau\approx 5$. 
Now waves 3 and 4 play the roles of the pumping waves, 
and wave 1 acts as a probing 
wave, wave 2 will become appreciable again until the pumping wave 4
vanishes at about $\tau\approx 10.5$. 
This completes one period of ideal 4WM.

Another special dynamical character in 4WM processes 
is the effect of relative phase difference.  
It is not the phase of each individual wave that matters.  
Instead if we write 
$\xi_{i} = \sqrt{I_i}e^{i\varphi_{i}}$, 
it is the combination of the phases 
$\alpha=\varphi_4+\varphi_3-\varphi_2-\varphi_1-\Omega\tau$
that affects 4WM in the way that
\begin{equation}
\label{eq:phase1}
\frac{\partial I_1}{\partial \tau}= 
\frac{\partial I_2}{\partial \tau}=
-\frac{\partial I_3}{\partial \tau}=
-\frac{\partial I_4}{\partial \tau}=
2{\mathcal{A}}\sin\alpha;
\end{equation}
\begin{equation}
\label{eq:phase2}
\frac{\partial \alpha}{\partial \tau}=-\Omega +{\mathcal{A}}
(\frac{1}{I_1}+\frac{1}{I_2}-\frac{1}{I_3}-\frac{1}{I_4})\cos\alpha,
\end{equation}
where the amplification factor 
is defined as ${\mathcal{A}}\equiv \sqrt{I_1I_2I_3I_4}$.
When one of the four waves vanishes, we can set the 
combined phase $\alpha$ to be an arbitrary value.  
Therefore the initial values 
of the pumping and probing waves do not affect the time evolution of atomic distributions.

\section{Conversion efficiency of 4WM}\label{sec:ce}
All of the previous works\cite{4wm1,4wm2,4wm3} on the 
4WM of matter waves have focussed on the initial rate of 
growth of the fourth wave because the time of wavepackets 
interaction in experiment\cite{deng} is very short 
(determined by the size and individual momentum of wavepackets).
Initially the rate of growth of the fourth wave 
is proportional to the product 
of the intensities of pumping and probing waves.
If the experimental setup can be changed in such a way 
that 4WM can proceed for a much longer period,
we need more precise knowledge about the conversion efficiency of 4WM.
 
In the experiment \cite{deng} of 4WM with matter waves in 
BEC with $N\approx 1.7\times 10^6$ sodium atoms
($M=3.848\times 10^{-23}g$ and $a_0=28\AA$), the  
pumping (1,2) and probing (3) waves are
prepared by Bragg scattering\cite{Bragg}.  It was estimated that  
the population numbers 
in the three states took on the initial values of  
$N_1^0\approx 6.8\times 10^5$, $N_2^0\approx 6.9\times 10^5$ and
$N_3^0\approx 3.3\times 10^5$, respectively. 
Assuming that 4WM took place in a region of volume 
${\mathcal{V}}=L^{3}$ with $L\approx 0.1 mm$, 
the characteristic time will be
$T\equiv ML^3/4Na_0h\approx 3ms $.  

The four curves in Fig. 4 show typical results of the 
conversion efficiency $N_4/N$ as a function of the 
re-scaled time $\tau$ obtained by Eqs.(\ref{eq:4WMa}) and (\ref{eq:4WMb}), 
hence assuming an ideal 4WM process with $\Omega=0$, 
for different sets of initial conditions.  
Curve (i) shows the situation in which 
$N_{1}^{0} = N_{2}^{0} \gg N_{3}^{0}$, which achieves a maximum efficiency 
of ~$50\%$ if 4WM can proceed for a time long enough, e.g., $t>10T$. 
Curve (ii) is calculated using a set of initial conditions 
similar to the experimental situation.
Comparing with curve (i), it demonstrates a much greater initial 
rate of growth of the fourth wave and a, however, smaller 
saturation efficiency.  
Note that a conversion efficiency of $10\%$ is achieved at $t \approx 1.9 T \approx 5.8 ms$,
which is consistent with the results reported in Ref.\cite{deng}.
The initial conditions for curves (iii) and (iv) are chosen 
so that both of them have the same initial rate of 
growth of the fourth wave.  Even so, the two curves demonstrate 
very different evolution processes as a function of $\tau$.
Therefore, carefully prepared initial atomic distributions
are crucial in achieving a higher initial rate of increase 
in the efficiency
and at the same time a high enough saturation efficiency.   

While the initial conditions in Ref.\cite{deng} are 
close to those values in achieving a high efficiency, the observed value of $6\sim 10\%$ is still rather low. 
A higher conversion efficiency can then 
be achieved by increasing the interaction time and/or by shortening 
the characteristic time by increasing the number density of atoms.

\section{Summary}\label{sec:summary}
In conclusion, a set of nonlinear equations for the time evolution of 
multi-wave mixing in BECs is dervied. 
The formalism leads to a  
physically transparent picture of 4WM with matter waves.
The ideal 4WM model is successfully applied to  analyse results 
of 4WM experiments. 
Factors affecting the value of the conversion efficiency are discussed.  
The GP equation in momentum space provides us with an efficient way to
study the wavepacket effects of 4WM because 
the four wavepackets are limited to four
spots of the momentum space even though 
they may spread widely in real space.

\section {Acknowledgments}\label{sec:thank}
This research is supported in part by the NSF of 
China for ``Climbing Project in Theoretical Physics", 
the Guangdong Provincial Natural Science Foundation of China,
and a Direct Grant for Research at the Chinese University of Hong Kong.

\newpage

\centerline{FIGURE CAPTIONS}

\begin{itemize}
\item[] Fig.1 \ \ \ The momentum sphere of phase-matching 
condition for 4WM.
The center is located 
at $({\mathbf{k}}_1+{\mathbf{k}}_2)/2$, and the radius of the sphere 
is defined as the magnitude of the 
vector $({\mathbf{k}}_1-{\mathbf{k}}_2)/2$.
If the momentum of the probing wave points exactly at a point on the 
surface of the sphere, the largest conversion 
efficiency of 4WM is achieved. 
\item[] Fig.2 \ \ \ The conversion coefficiency  
as a function of interaction time for five different values of the 
phase-mismatch paramter $\Omega = T \Delta \omega$. 
The time is given in units of the characteristic time as defined 
in the text. 
The five curves from top to bottom correspond to 
$\Omega/\pi=0,\;0.1,\;0.2,\;0.4,\;1$, respectively. 
The initial conditions for the pumping and probing waves 
are chosen to be $\xi_1^0,\; \xi_2^0,\; \xi_3^0= 0.68,\; 0.58,\; 0.45$.
\item[] Fig.3 \ \ \ Typical time evolution curves of 
the populations of the matter waves involved in a 4WM process. 
  The initial conditions are chosen 
to be $\xi_1^0,\; \xi_2^0,\; \xi_3^0= 0.68,\; 0.58,\; 0.45$.  
\item[] Fig.4 \ \ \ Typical results of the 
conversion efficiency as a function of the interaction time for 
different initial conditions. 
The initial conditions for curve (i) are 
$\xi_1^0,\; \xi_2^0,\; \xi_3^0= 0.707,\; 0.703,\; 0.243$, which 
lead to a maximum efficiency of $~50\%$ but with a low initial 
increasing rate. 
Curve (ii) has initial conditions similar to those 
in experimental situation: 
$\xi_1^0,\; \xi_2^0,\; \xi_3^0= 0.637,\; 0.632,\; 0.441$.  
Curves (iii) and (iv) 
show that in order to achieve an appreciable efficiency, 
the probing wave can be weak but the pumping waves must be strong. 
The initial conditions are: (iii)
$\xi_1^0,\; \xi_2^0,\; \xi_3^0= 0.804,\; 0.553,\; 0.217$
and (iv) $\xi_1^0,\; \xi_2^0,\; \xi_3^0= 0.804,\; 0.217,\; 0.553$.
\end{itemize}

\end{document}